# Compositionally Complex Ceramics (CCCs): A New Paradigm in Ceramics

Jian Luo [*]

Aiiso Yufeng Li Family Department of Chemical and Nano Engineering; Program in Materials Science and Engineering, University of California San Diego, La Jolla, California 92093, U.S.A.

## Abstract

The development of high-entropy ceramics (HECs) over the past decade has extended the exploration to a diverse range of oxides, borides, silicides, carbides, nitrides, fluorides, and other ceramic solid solutions, encompassing an increasingly broad range of crystal structures and bonding characteristics and exhibiting promising mechanical, thermal, and functional properties. Initial studies predominantly focused on five-component equimolar compositions, often assuming the formation of random solid solutions. In 2020, we proposed extending the exploration beyond HECs to the broader class of "compositionally complex ceramics" (CCCs), in which rational non-equimolar compositional designs and long- and short-range order can reduce configurational entropy while providing additional opportunities to tailor and enhance materials properties, with the goal of outperforming their higher-entropy counterparts. Because CCCs have been defined somewhat vaguely in previous publications, this review seeks to establish more rigorous and explicit criteria for defining CCCs, despite the unavoidable degree of arbitrariness in the choice of thresholds. Meanwhile, 10–21-component ultrahigh-entropy ceramics have emerged as a subset of HECs, exhibiting intriguing abrupt phase transitions. Dual-phase CCCs have also been reported, with thermodynamic equilibria governing cation partitioning between the two phases and offering additional opportunities to tailor properties through microstructural engineering. Recent studies have further revealed that grain-boundary complexion (phase-like) transitions can occur in CCCs, influencing microstructural evolution and materials properties. Overall, CCCs provide a versatile platform for tailoring materials properties through the interplay of diverse crystal structures and bonding characteristics, compositional complexity, rational non-equimolar designs, long- and short-range order, defects, and microstructural and interfacial engineering.

## Graphical Abstract

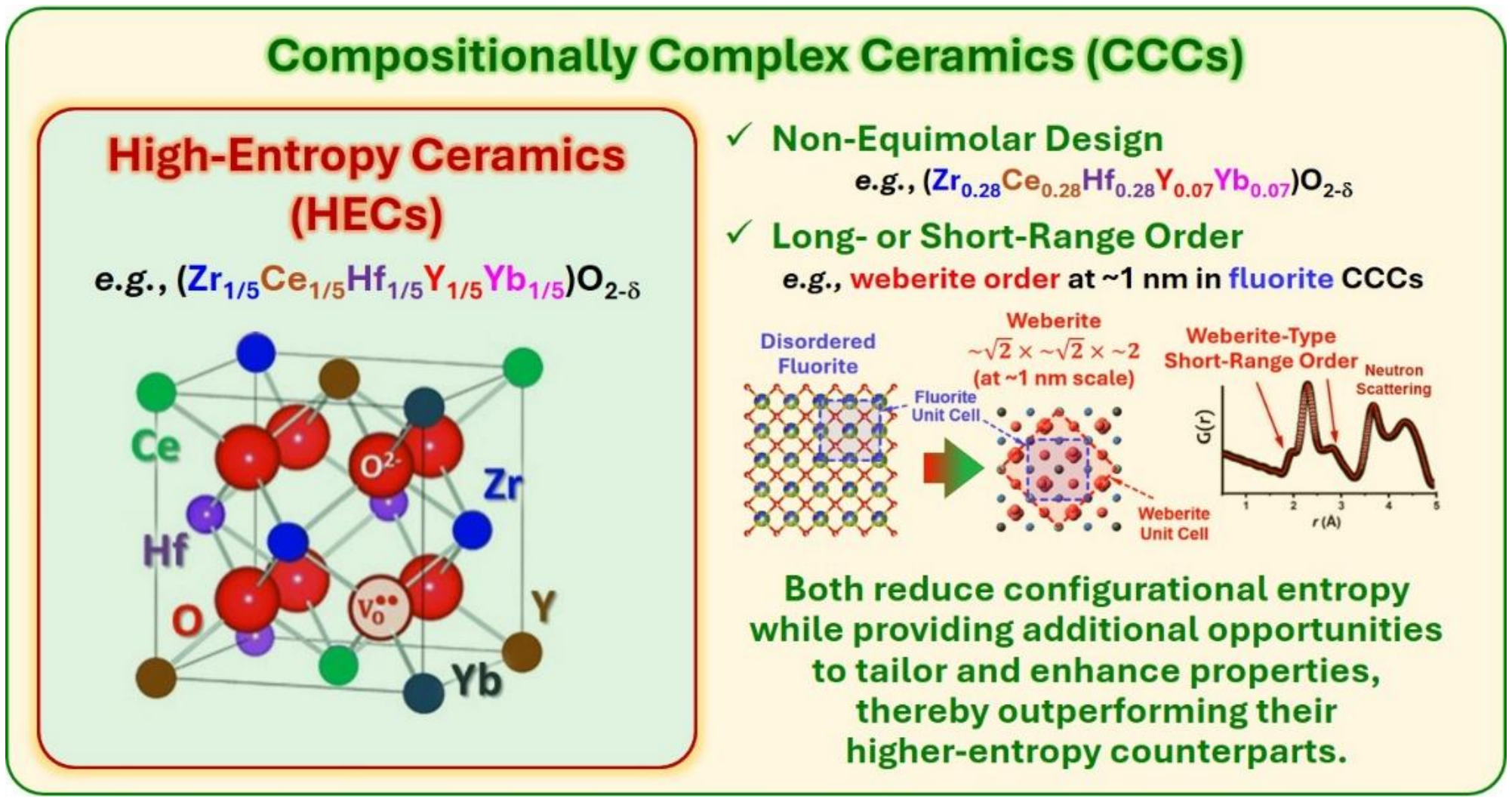

[*] Email: jluo@alum.mit.edu

## Introduction: High-Entropy Ceramics (HECs)

Inspired by research on high-entropy alloys (HEAs), also known as multi-principal-element alloys (MEPAs) or complex concentrated alloys (CCAs), the field of high-entropy ceramics (HECs) has expanded rapidly over the past decade to encompass a wide range of oxides, borides, silicides, carbides, nitrides, fluorides, silicates, and other ceramic systems, initially focusing on equimolar quinary solid solutions.[1, 2] Figure 1 shows selected representative HECs.

Following our earlier perspective articles, which discuss relevant terminologies and definitions,[1, 2] here I adopt a somewhat arbitrary definition of *high-entropy ceramics* (*HECs*) as ceramic solid solutions with an ideal mixing configurational entropy greater than 1.5 $k_B$ per atom (or per cation in ionic ceramics) on at least one sublattice (or Wyckoff position):

$$\max_i\{\Delta S^{\text{ideal mix}}_{\text{per atom on Sublattice } i}\} > 1.5\ k_B \tag{1}$$

where $k_B$ is the Boltzmann constant. Similarly, *medium-entropy ceramics* (*MECs*) are defined as ceramic solid solutions with the highest ideal mixing configurational entropy on any sublattice between 1 and 1.5 $k_B$ per atom, whereas *low-entropy ceramics* (*LECs*) have an ideal mixing configurational entropy below 1 $k_B$ per atom on all sublattices. It is worth noting that "per cation" is often used, particularly for oxides, because mixing typically occurs on a cation-like metal sublattice. However, some non-oxide HECs have low ionicity. For example, in the intermetallic $(Fe_{1/5}Co_{1/5}Ni_{1/5}Mn_{1/5}Cu_{1/5})Al$ (Figure 1l),[3] mixing occurs on the anion-like sublattice (despite its very low ionicity and predominantly metallic bonding character). Importantly, these cutoff values, analogous to those used for HEAs[4] but applied to ceramic sublattices, are subjective and may not be scientifically or practically important. Moreover, these definitions are based on *ideal configurational entropy* and therefore do not account for short-range order, which can reduce the actual configurational entropy but is often difficult to quantify.

The field of metallic HEAs was pioneered by two seminal reports by Yeh *et al.*[5] and Cantor *et al.*[6] in 2004. Subsequently, high-entropy nitride,[7] oxide,[8] and carbide[9] coatings or thin films were first reported in 2006,[7] 2007,[8] and 2012,[9] respectively. In 2015, Rost *et al.* reported an entropy-stabilized oxide, $(Co_{1/5}Cu_{1/5}Mg_{1/5}Ni_{1/5}Zn_{1/5})O$, in the rocksalt structure (Figure 1a),[10] stimulating interest in fabricating HECs in bulk form.

In 2016, Gild *et al.* reported high-entropy diborides (*e.g.*, $(Ti_{1/5}Zr_{1/5}Hf_{1/5}Nb_{1/5}Ta_{1/5})B_2$; Figure 1e)[11] as a new class of ultrahigh-temperature ceramics (UHTCs) and the first high-entropy borides. Together with subsequent reports of high-entropy rocksalt carbides[12-14] and nitrides[15] with similar cation combinations (*e.g.*, $(Ti_{1/5}Zr_{1/5}Hf_{1/5}Nb_{1/5}Ta_{1/5})C$ and $(Ti_{1/5}Zr_{1/5}Hf_{1/5}Nb_{1/5}Ta_{1/5})N$; Figure 1i), this work[11] inspired numerous follow-up studies of high-entropy UHTCs worldwide for extreme-environment applications.[16-19] Notably, the metallurgy community had already demonstrated that refractory HEAs such as VNbMoTaW can retain high strength at elevated temperatures (Figure 2a).[20-23] An even more intriguing opportunity is suggested by the exceptional flexural strength of $(Ti_{1/5}Zr_{1/5}Hf_{1/5}Nb_{1/5}Ta_{1/5})B_2$ shown in Figure 2b, which surprisingly increases with temperature to >700 MPa at 1800–2000 °C, more than three times that of the state-of-the-art UHTC $ZrB_2$.[24, 25] Various mechanisms may contribute to this ultrahigh-temperature strengthening, motivating further fundamental studies of the underlying deformation mechanisms in future. In addition, a range of high-entropy borides with different stoichiometries, crystal structures, and bonding characteristics have also been reported, including high-entropy monoborides (Figure 1h),[26] the first reported high-entropy superhard materials in 2020, followed by high-entropy $M_3B_4$ (Figure 1g) and $MB_4$ (Figure 1f) borides in 2021.[27, 28]

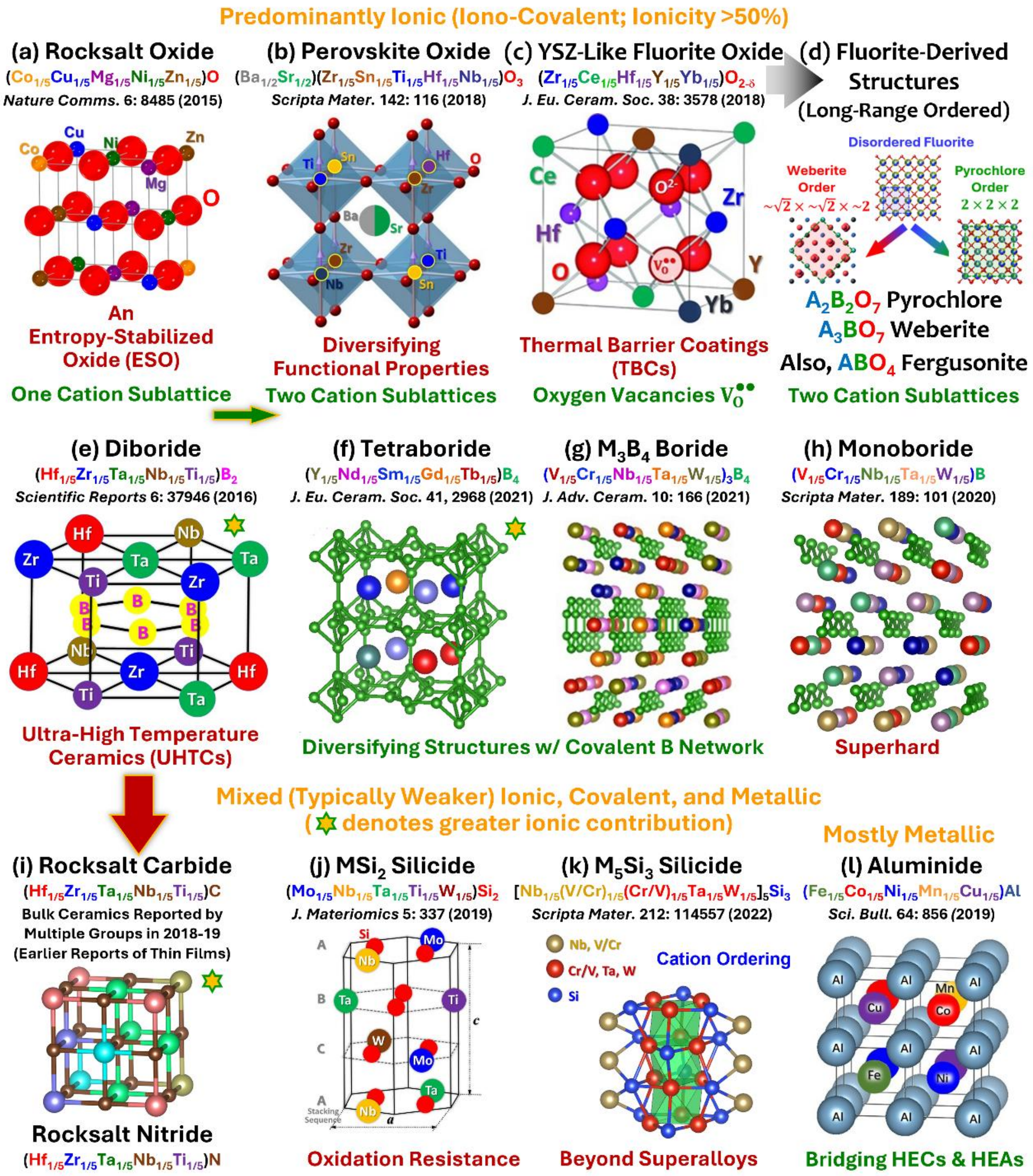

**Figure 1.** Selected representative high-entropy ceramics (HECs), including (a) a rocksalt entropy-stabilized oxide (ESO);[10] (b) a high-entropy $ABO_3$ perovskite oxide,[29] the first HEC with two cation sublattices and a broad spectrum of functional properties; (c) a YSZ-like high-entropy fluorite oxide,[30] which stimulated interest in exploring a new class of high-entropy thermal barrier coating (TBC) materials; (d) fluorite-derived high-entropy pyrochlore, weberite, and fergusonite oxides;[31, 32] (e) a high-entropy $MB_2$ boride,[11] the first high-entropy ultrahigh-temperature ceramics (UHTCs);[11] high-entropy (f) $MB_4$ and $M_3B_4$ (g) borides;[27, 28] (h) a high-entropy MB boride,[26] the first high-entropy superhard materials;[26] (i) high-entropy rocksalt carbides[12-14] and nitrides[15], also high-entropy UHTCs with similar metal/cation combinations as high-entropy metal diborides; (j) a high-entropy $MSi_2$ silicide,[33] the first high-entropy silicide with superior intermediate-temperature oxidation resistance; (k) a quinary $M_3Si_5$ silicide with unusual phase stability and cation ordering,[34] which may enable a new class of high-temperature composites beyond superalloys; and (l) a high-entropy aluminide,[3] the first reported single-phase high-entropy intermetallic compound (with mostly metallic bonding) bridging HECs and high-entropy alloys (HEAs). Schematics are adapted from References 1, 10, 11, 33, 34 under CC BY open-access licenses and reproduced with permission from References 3, 26-29, 35.

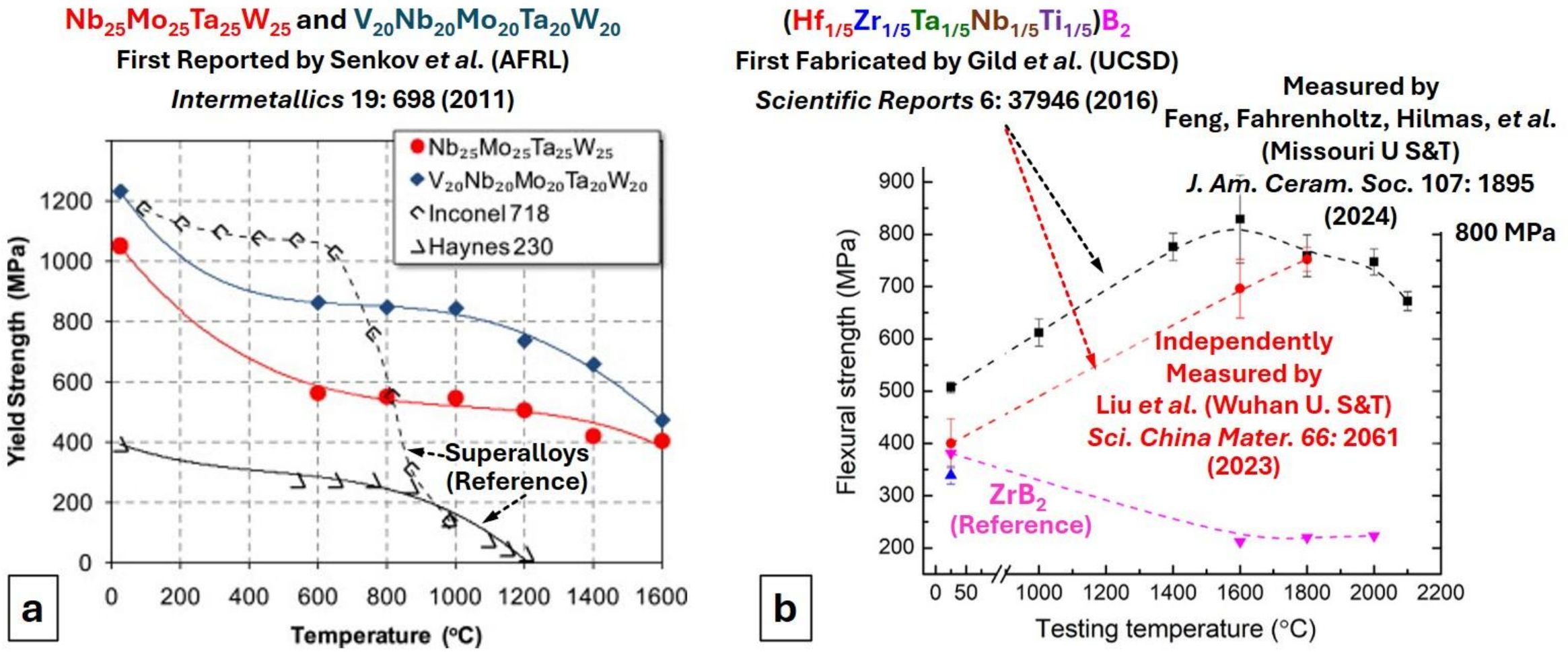

**Figure 2.** Comparison of high-temperature strengths of refractory high-entropy alloys and ceramics. (a) Temperature dependence of the yield stress of refractory high-entropy alloys reported by Senkov *et al.*,[23] compared with those of two reference Ni-based superalloys, Inconel 718 and Haynes 230. (b) Two independent studies showed that the flexural strength of $(Zr_{1/5}Hf_{1/5}Ti_{1/5}Ta_{1/5}Nb_{1/5})B_2$ increases with increasing temperature and exceeds that of the $ZrB_2$ benchmark by more than threefold at 1800–2000 °C.[24, 25] Reprinted with permission from References 23, 24.

In 2018, Jiang *et al.* reported the first high-entropy $ABO_3$ perovskite oxides (Figure 1b) with two cation sublattices and diverse functional properties,[29] stimulating worldwide interest in exploring their catalytic,[36-38] dielectric,[39-44] ferroelectric,[45-57] magnetic,[58-61] thermoelectric,[62-64] magnetocaloric,[65] electrocaloric,[44] and quantum properties,[66] as well as their potential applications in fuel cells,[67-72] batteries,[73-76] supercapacitors,[77] hydrogen generation,[78-80] and other areas.[55, 81-84]

In 2018, Gild *et al.* reported yttria-stabilized zirconia (YSZ)-like high-entropy fluorite oxides with aliovalent doping and oxygen vacancies as a new class of thermal-barrier-coating (TBC) materials (Figure 1c).[30] Subsequent studies investigated a series of medium- and high-entropy oxides with long-range order in fluorite and fluorite-derived pyrochlore,[32, 85] weberite,[31] and fergusonite[31] structures (Figure 1d), along with possible short-range order and order–disorder transitions,[31, 32, 86-90] offering diverse opportunities to tailor thermal and mechanical properties.

In 2019, Gild *et al.* also reported the first high-entropy silicide, $(Mo_{1/5}Nb_{1/5}Ta_{1/5}Ti_{1/5}W_{1/5})Si_2$ (Figure 1j),[33] along with an independent report of a slightly different composition published at approximately the same time.[91] A follow-up study reported the intriguing, and highly desirable, "complete elimination of pest oxidation" in high-entropy disilicides.[92] In 2022, Shivakumar *et al.* reported a second class of quinary $M_5Si_3$ silicides with unusual polytype formation and cation ordering.[34] Notably, $(V_{1/5}Cr_{1/5}Nb_{1/5}Ta_{1/5}W_{1/5})_5Si_3$ (Figure 1k) forms the hexagonal γ ($D8_8$) phase with cation ordering, despite all five constituent binary silicides, $V_5Si_3$, $Cr_5Si_3$, $Nb_5Si_3$, $Ta_5Si_3$, and $W_5Si_3$, being stable in the tetragonal α ($D8_l$) or β ($D8_m$) phases.[34] Although cation ordering likely reduces $(V_{1/5}Cr_{1/5}Nb_{1/5}Ta_{1/5}W_{1/5})_5Si_3$ to a MEC, formation of the hexagonal γ phase is favorable for high-temperature mechanical properties, potentially enabling a new class of high-entropy analogs of Ni-silicide composites for applications beyond superalloys.

In contrast to the metallic bonding in HEAs, oxide HECs are predominantly ionic (iono-covalent, often with >50% ionicity) due to the high electronegativity of oxygen. Boride, silicide, carbide, and nitride HECs can exhibit complex combinations of metallic, covalent, and typically weaker ionic bonding in varying proportions. Notably, the ionic contribution is greater in diborides, rare earth tetraborides, and rocksalt-structured carbides and nitrides, where localized charge transfer between distinct metal and non-metal sublattices is enhanced. In 2019, Zhou *et al.* reported a single-phase high-entropy intermetallic compound, $(Fe_{1/5}Co_{1/5}Ni_{1/5}Mn_{1/5}Cu_{1/5})Al$ (Figure 11),[3] further providing a link between HEAs and HECs by combining high-entropy mixing confined to a single (anion-like) sublattice with predominantly metallic bonding character, albeit modified by directional covalent hybridization. Ultimately, the diverse chemistries, crystal structures, and bonding characters of HECs enable a broad spectrum of structural and functional properties.

*Is the entropy in HECs truly high?* The contribution of configurational entropy to Gibbs free energy is comparable to the thermal vibrational energy (1.5 $k_B T$ per atom) and is generally modest, particularly when compared with the large enthalpy changes associated with reactions such as oxidation in non-oxide HECs. Nevertheless, it can be sufficient to stabilize a single-phase HEC in some cases, as demonstrated by $(Co_{1/5}Cu_{1/5}Mg_{1/5}Ni_{1/5}Zn_{1/5})O$.[10] See a recent Perspective for further discussion.[1]

*Should we always maximize entropy to achieve desired properties?* The answer is likely no, as discussed below.

## From HECs to Compositionally Complex Ceramics (CCCs)

In a 2020 study[93] and a subsequent perspective article,[2] we proposed broadening the research on *high-entropy ceramics* (*HECs*) to *compositionally complex ceramics* (*CCCs*) (Figure 3), emphasizing non-equimolar compositional designs and long- or short-range order, both of which reduce configurational entropy while providing additional opportunities to tailor and enhance materials properties, thereby outperforming their higher-entropy counterparts.[1, 2]

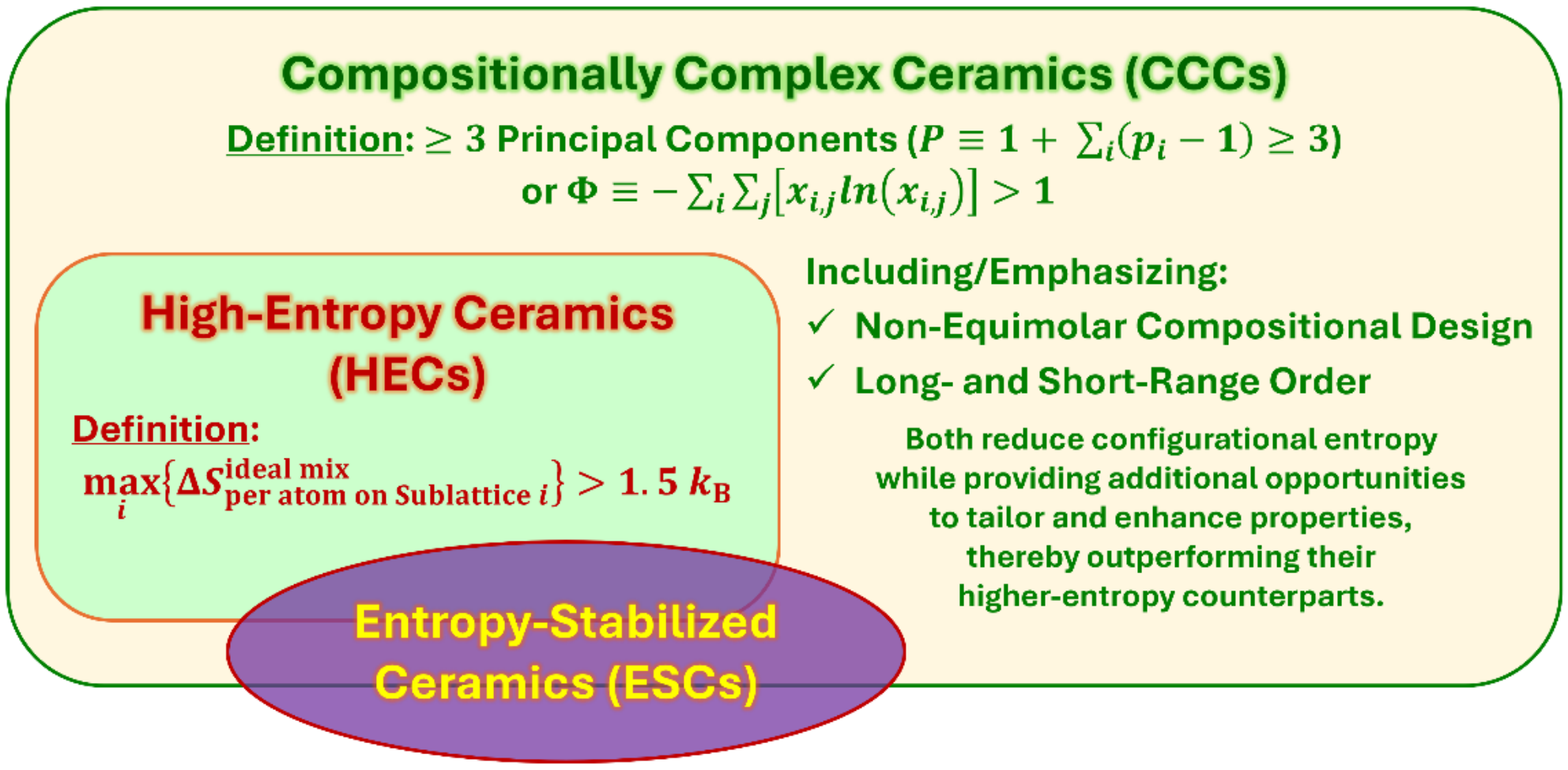

**Figure 3.** From high-entropy ceramics (HECs) to compositionally complex ceramics (CCCs). HECs are defined as ceramic solid solutions with >1.5 $k_B$ per atom (or per cation) of ideal configurational entropy of mixing on at least one sublattice. In a 2020 study[93] and a subsequent perspective article,[2] we proposed broadening the scope of HECs to CCCs, in which non-equimolar compositions and long- or short-range order provide additional opportunities to tailor and enhance properties despite reduced configurational entropy, thereby outperforming their higher-entropy counterparts. Here, I broadly define CCCs as ceramic solid solutions comprising at least three principal components ($P \geq 3$, as defined in Equation 3) or satisfying the condition $\Phi > 1$ (as defined in Equation 2). Entropy-stabilized ceramics (ESCs) can be HECs, CCCs, or other simpler (including low-entropy) ceramics, as discussed in the text. This figure is revised from an original plot in Reference 93.

Here, I broadly define "compositionally complex ceramics" (CCCs) as ceramic solid solutions comprising at least three principal components ($P \geq 3$) or satisfying the following condition:

$$\Phi \equiv -\sum_i \sum_j [x_{i,j} ln(x_{i,j})] > 1, \tag{2}$$

where $x_{i,j} \equiv x_{\text{Element } j}^{\text{Sublattice } i}$ is the fraction of element $j$ on Sublattice $i$, and $\Phi$ is a variable introduced to quantify the compositional complexity of a solid solution involving mixing on multiple sublattices. This definition comprises two conditions, either of which qualifies a ceramic solid solution as a CCC. These two conditions have substantial overlaps, encompassing the majority of CCCs under both definitions.

The first condition, requiring at least three principal components, has been used previously.[1, 2] Here, however, a principal component needs to be defined. In the field of metallic HEAs (or MPEAs and CCAs), a principal element is generally considered to have an atomic fraction between 5% and 35%,[4] a somewhat arbitrary range. I alternatively proposed defining a principal component as one whose molar fraction is no less than one-third that of the most concentrated component on the corresponding sublattice.[1] To account for the compositional complexity arising from mixing on multiple sublattices, here I further propose to define the number of principal components, $P$, for a solid solution containing one or more principal element(s) on each of $N$ sublattices ($N \geq 1$) as:

$$P \equiv 1 + \sum_{i=1}^{N} (p_i - 1), \tag{3}$$

where $p_i$ is the number of principal elements on the $i^{\text{th}}$ sublattice. The above equation indicates that $p_i$ principal elements on the $i^{\text{th}}$ sublattice will contribute $(p_i - 1)$ independent compositional variable(s), which can be considered additional principal component(s). In this definition, the number of independent principal components is always $P - 1$, regardless of the number of sublattices. In other words, the first condition requires at least two independent principal components ($P - 1 \geq 2$) beyond the base compound, which may itself contain multiple elements but is counted as a single component.

Equation (2) means that the sum of the ideal configurational entropies of mixing across all sublattices, normalized on a per-atom basis for each sublattice, exceeds 1 $k_B$: $\sum_i \Delta s_{\text{per atom on Sublattice } i}^{\text{ideal mix}} > 1\ k_B$. For example, for a perovskite $ABO_3$ with mixing on both the A and B sublattices, $\Phi = ln\Omega$, where $\Omega$ is the configurational multiplicity of the solid solution, and $\Phi \cdot R$ corresponds to its ideal configurational entropy per mole of $ABO_3$. However, this relationship does not hold for weberite $AB_3O_7$, where mixing on the B sublattice contributes three times to the configurational multiplicity and ideal configurational entropy because there are three B-site cations per formula unit. Nevertheless, the compositional complexity should not be considered three times greater. In other words, configurational entropy accounts for the number of mixing sites per formula unit, whereas $\Phi$ is intended to characterize compositional complexity independently of the multiplicity of a particular sublattice.

The second condition (Equation 2) is introduced to broaden the definition of CCCs for two reasons: (1) to ensure that all HECs and MECs are encompassed as subsets of CCCs; and (2) to quantitatively account for the compositional complexity arising from mixing across multiple sublattices, thereby providing a more quantitative measure of compositional complexity than Equation (3). This newly proposed second condition for defining CCCs is generally broader than, but does not fully encompass, the (revised) first condition.

Multiple definitions of CCCs can be proposed based on either ideal configurational entropy or the number of principal components.[1, 2] Although different definitions generally encompass similar classes of materials, the specific range of CCCs depends on the definition adopted. For example, an earlier, simpler definition considered CCCs as the combination of MECs and HECs (CCCs = MECs + HECs),[2] whereas the current definition encompasses a broader range of compositions (CCCs ⊃ MECs + HECs), emphasizing compositional complexity (e.g., $(Ba_{0.5}Sr_{0.5})(Ti_{0.5}Zr_{0.5})O_3$ and $(Zr_{0.55}Hf_{0.27}Ti_{0.18})B_2$ qualifies as CCCs but are LECs under current definitions).

As with HECs and MECs, definitions of CCCs are inherently subjective, and the threshold values used to define CCCs may not necessarily have clear scientific or technological significance and are therefore somewhat arbitrary. Thus, which particular compositions qualify as CCCs may be less important than the underlying design philosophy. Rather than pursuing maximum configurational entropy, the design of CCCs focuses on leveraging compositional complexity, non-equimolar compositions, and long- and short-range order to tailor and enhance materials properties, with the goal of outperforming their higher-entropy counterparts.[1, 2]

It is also worth discussing the concept of "entropy-stabilized ceramics" (ESCs). In a broad definition, if an enthalpic penalty is overcome by an entropic gain to stabilize a phase, that phase is considered entropy stabilized. Based on this definition and thermodynamic principles, if a temperature-induced phase transformation produces a single ceramic phase at high temperature, that high-temperature phase can be considered an ESC. The concept of entropy stabilization in oxides was introduced by Navrotsky in the late 1960s during the thermodynamic analysis of binary spinel solid solutions that are LECs.[94] Rost *et al.* proposed a more restrictive set of criteria for multicomponent (high-entropy) entropy-stabilized oxides.[10] It should be noted that many HECs and CCCs are not ESCs, and not all ESCs are HECs or CCCs (Figure 3).

## Non-Equimolar Designs

One of the key concepts in CCC design is the exploration of non-equimolar compositions, with the goal of outperforming their higher-entropy, equimolar counterparts. In 2020, Wright *et al.* first demonstrated the benefits of non-equimolar design in CCCs by investigating the thermal conductivity ($k$) and elastic modulus ($E$) of a series of YSZ-like fluorite oxides, $(Hf_{1/3}Zr_{1/3}Ce_{1/3})_{1-x}(Y_{1/2}Yb_{1/2})_xO_{2-\delta}$ (Figure 4).[93] Conventionally, low thermal conductivity is associated with "soft" materials. A distinct feature of CCCs, however, is reduced thermal conductivity accompanied by retained elastic modulus, thereby breaking the traditional trade-off between $k$ and $E$. Thus, CCCs can simultaneously achieve low $k$ and high $E$, resulting in an increased $E/k$ ratio. Specifically, the non-equimolar composition $(Hf_{0.28}Zr_{0.28}Ce_{0.28}Y_{0.07}Yb_{0.07})_xO_{2-\delta}$ achieved a higher $E/k$ ratio and demonstrated superior potential as a TBC material, outperforming its higher-entropy equimolar counterpart, $(Hf_{1/5}Zr_{1/5}Ce_{1/5}Y_{1/5}Yb_{1/5})_xO_{2-\delta}$ (Figure 4).[93] In this case, when $x$ exceeds 0.07, the formation of more than 5% oxygen vacancies in the anion sublattice would likely promote vacancy clustering and ordering, thereby suppressing phonon scattering, increasing $k$, and reducing the $E/k$ ratio (Figure 4).

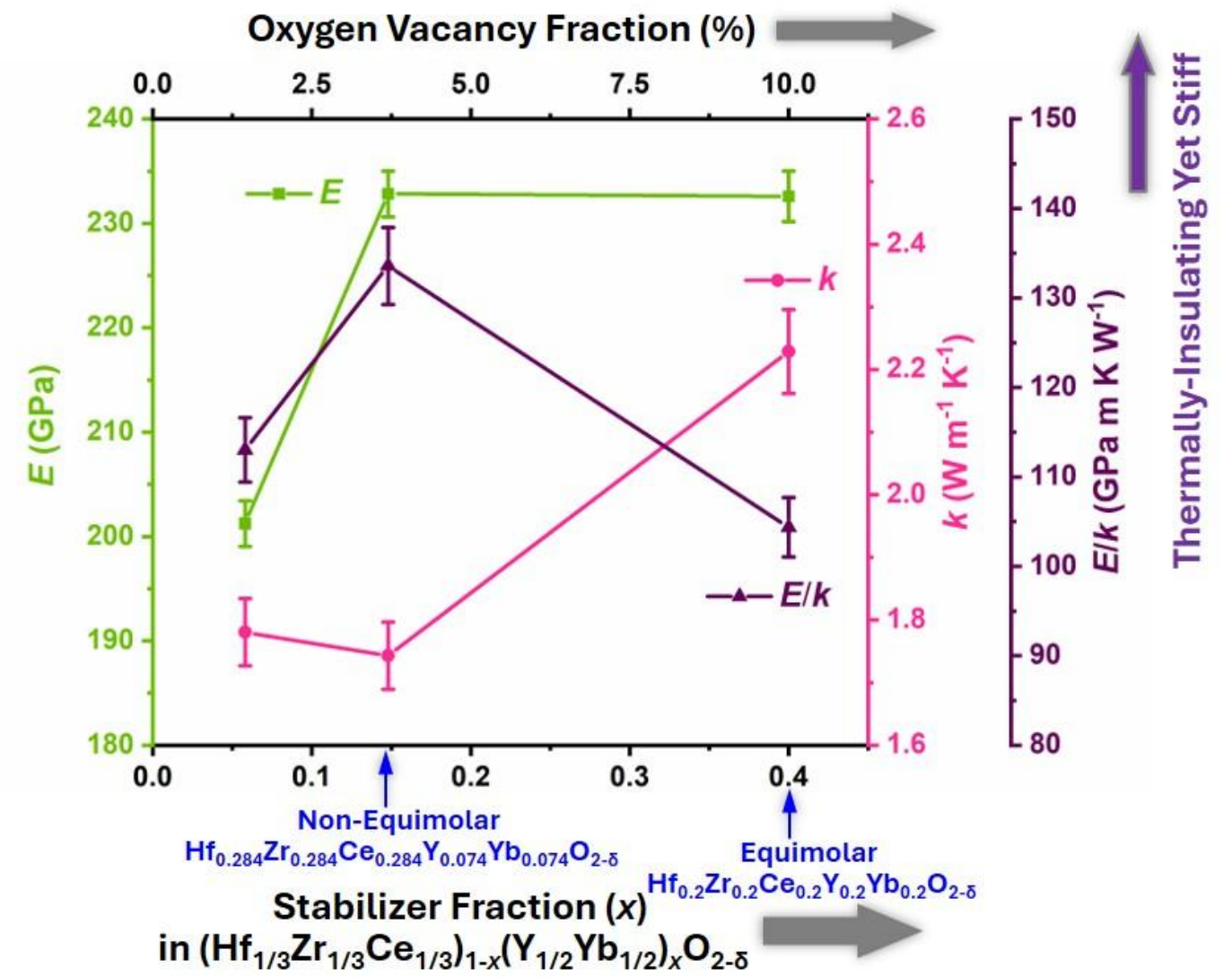

**Figure 4.** A non-equimolar composition, $(Hf_{0.28}Zr_{0.28}Ce_{0.28}Y_{0.07}Yb_{0.07})_xO_{2-\delta}$, achieves superior thermomechanical properties for thermal-barrier-coating (TBC) applications, outperforming the higher-entropy equimolar composition $(Hf_{1/5}Zr_{1/5}Ce_{1/5}Y_{1/5}Yb_{1/5})_xO_{2-\delta}$.[93] Shown are the measured Young's modulus ($E$), thermal conductivity ($k$), and $E/k$ ratio

for three $(Hf_{1/3}Zr_{1/3}Ce_{1/3})_{1-x}(Y_{1/2}Yb_{1/2})_xO_{2-\delta}$ compositions. The non-equimolar $(Hf_{0.28}Zr_{0.28}Ce_{0.28}Y_{0.07}Yb_{0.07})_xO_{2-\delta}$ exhibits the lowest thermal conductivity and highest $E/k$ ratio. Replotted with permission from References 2, 93.

In some cases, non-equimolar designs in CCCs are necessary for phase stability because of charge balance, such as with aliovalent substitution on one or more sublattices, as well as other structural criteria related to the average cation size on a sublattice, such as satisfying the Goldschmidt tolerance factor criterion in perovskite oxides. Several examples are discussed subsequently.

Exploring non-equimolar compositions in CCCs is nontrivial because it introduces vastly larger compositional design spaces. By leveraging compositional complexity and non-equimolar designs, CCCs offer greater opportunities for designing ceramics with exceptional functionalities.

## Long- and Short-Range Order and Correlated Disorder

Long- and short-range order, which also reduce configurational entropy, can exist in CCCs and be leveraged to tailor and enhance properties. For example, cubic defect fluorite oxides can exhibit pyrochlore (cubic $2 \times 2 \times 2$ superstructure) and weberite-type (e.g., orthorhombic $\sim\sqrt{2} \times \sim\sqrt{2} \times \sim 2$ superstructure in $C222_1$ local symmetry) order, involving cation and oxygen-vacancy ordering (Figure 1d), which can influence thermal and mechanical properties.[31, 86-90]

In addition to long-range order, various types of short-range order can exist and affect a range of properties. In the metallurgy community, chemical short-range order (CSRO) in HEAs has attracted significant recent interest because of its potential influence on mechanical properties.[4] In addition to CSRO, structural short-range order has also been observed in CCCs. For example, two neutron total scattering studies revealed that CCCs with a long-range cubic fluorite structure can exhibit orthorhombic weberite-type short-range order at the ~1 nm scale, which can contribute to ultralow thermal conductivity.[86, 88]

Wright *et al.*[86] discovered a negative correlation between thermal conductivity and the average radius ratio of the 3+/5+ cations in rare-earth niobates and tantalates, which adopt a long-range disordered defect-fluorite structure. Because the radius ratio remains below the threshold for forming long-range-ordered weberite-type phases, while long-range-ordered weberite oxides are expected to have higher thermal conductivity, this correlation suggests that the ultralow thermal conductivity originates from short-range rather than long-range weberite-type order. Neutron total scattering was used to characterize short-range order in five CCCs, $(Sc_{0.25}Yb_{0.25}Lu_{0.25}Nb_{0.25})_4O_7$, $(Sc_{0.165}Dy_{0.191}Ho_{0.197}Tm_{0.197}Nb_{0.25})_4O_7$, $(Er_{0.25}Tm_{0.25}Yb_{0.25}Nb_{0.25})_4O_7$, $(Dy_{0.25}Ho_{0.25}Er_{0.25}Nb_{0.125}Ta_{0.125})_4O_7$, and $(Sc_{0.25}Yb_{0.25}Lu_{0.25}Nb_{0.25})_4O_7$. Small-box modeling of the neutron total scattering data revealed a weberite-type short-range order at the ~1 nm scale, with the domain size appearing to decrease in the more insulating compositions.[86]

Such short-range order can be considered *correlated disorder*, which can give rise to local symmetries that are distinct from, and typically lower than, the global symmetry.[95] Here, I hypothesize that such correlated disorder can be amplified by the compositional complexity of CCCs, leading to the formation of nanoscale domains with coupled chemical and structural short-range order that can be harnessed to achieve a range of extraordinary mechanical, thermal, and functional properties. For example, I hypothesize that correlated disorder, or the formation of nanoscale domains with coupled chemical and structural short-range order, contributes to the ultralow thermal conductivity of fluorite-based oxide CCCs with ~1 nm weberite-type short-range order,[86, 88] the unusual hardening of $(Zr_{1/5}Hf_{1/5}Ti_{1/5}Ta_{1/5}W_{1/5})B_2$ upon adding softer $WB_2$,[96] and the ultrahigh dielectric energy-storage densities of perovskite-based CCCs as ferroelectric relaxors.[53]

Figure 5 illustrates long- and short-range order in a series of single-phase, 10-component CCCs, $[(Pr_{0.375}Nd_{0.375}Yb_{0.25})_2(Ti_{0.5}Hf_{0.25}Zr_{0.25})_2O_7]_{1-x}[(DyHoErNb)O_7]_x$ $(0 \leq x \leq 1)$, denoted $10CCFBO_{xNb}$.[88] A long-range order–disorder transition (ODT) occurs at $x = 0.81 \pm 0.01$, from ordered pyrochlore to disordered defect fluorite (Figure 5a). In contrast to ternary oxides, this ODT occurs abruptly without an observable two- or multi-phase region. Rietveld refinements of neutron diffraction patterns suggest that this ODT involves migration of oxygen anions from the 48*f* to 8*a* sites, with a small final jump at the ODT, although the 8*a* oxygen occupancy changes gradually (Figure 5b). Diffuse scattering in Nb-rich compositions further indicates the presence of short-range order (Figure 5c). Small-box modeling shows that four compositions near the ODT ($x$ = 0.75, 0.8, 0.85, and 1) are better described by weberite-type ordering of the local polyhedral structure at the nanoscale. Interestingly, $10CCFBO_{0.75Nb}$ and $10CCFBO_{0.8Nb}$ exhibit both long-range pyrochlore order and short-range weberite-type order. Thus, weberite-type short-range order emerges before the long-range ODT, coexisting with long-range pyrochlore order. After the ODT, the long-range pyrochlore order disappears, while the short-range weberite-type order persists in the fluorite structure. Notably, the decrease in thermal conductivity coincides with the emergence of short-range order rather than the long-range ODT.

A further study[90] demonstrated that a reversible fluorite–pyrochlore ODT can be induced in $10CCFBO_{0.8Nb}$: $(Nd_{0.15}Pr_{0.15}Dy_{0.8}Ho_{0.8}Er_{0.8}Ti_{0.2}Yb_{0.1}Hf_{0.1}Zr_{0.1}Nb_{0.8})O_{7-\delta}$, by annealing at 1600 °C in oxidizing versus reducing environments. Notably, this 10-cation CCC remains a homogeneous, single-phase solid solution before and after the ODT, adopting pyrochlore and fluorite structures, respectively. *In situ* neutron diffraction revealed oxygen-vacancy formation and atomic displacements during the ODT. This study demonstrates a new pathway for inducing ODTs through redox transitions to tailor the properties of CCCs.

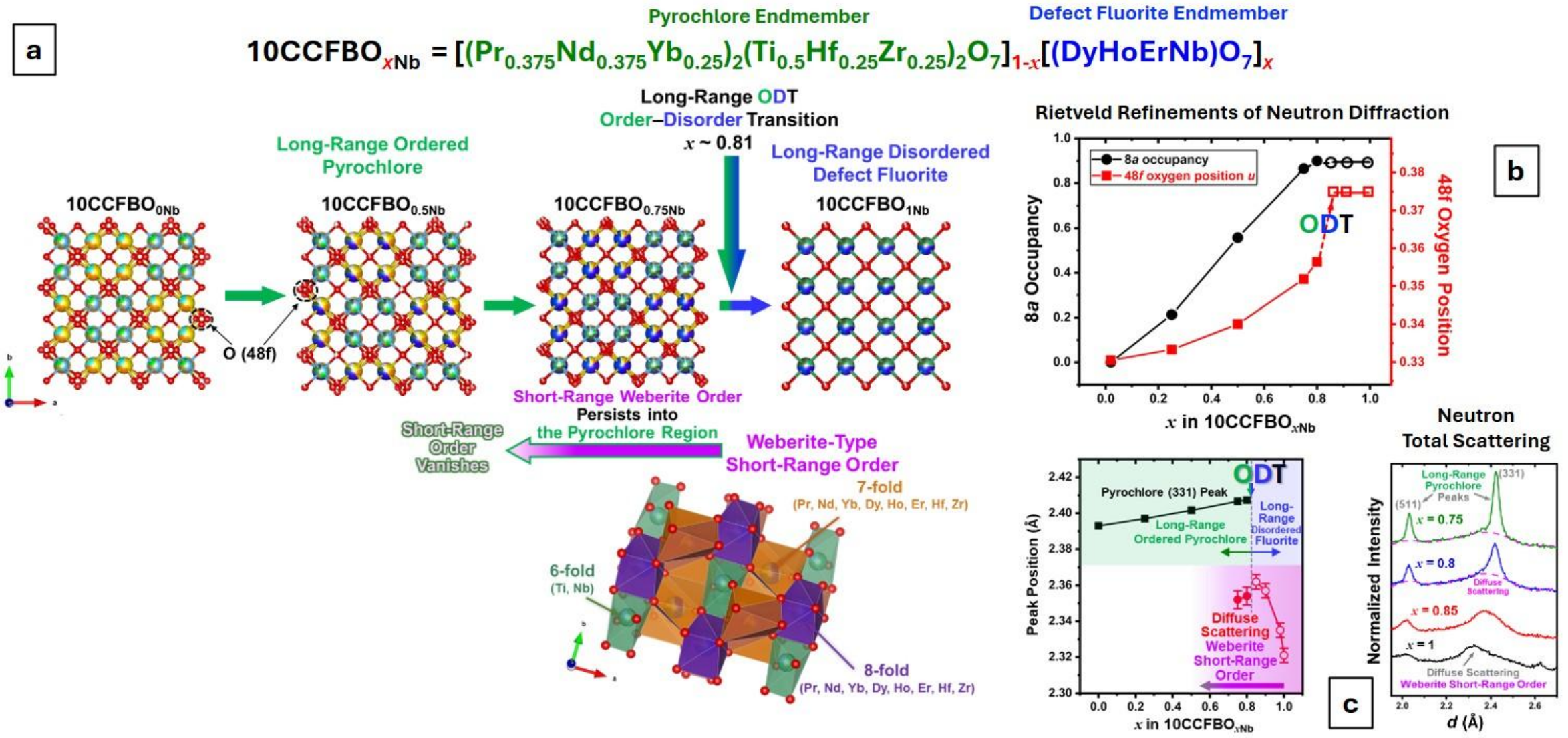

**Figure 5.** Long- and short-range order in a series of single-phase, 10-component, compositionally complex fluorite-based oxides, $[(Pr_{0.375}Nd_{0.375}Yb_{0.25})_2(Ti_{0.5}Hf_{0.25}Zr_{0.25})_2O_7]_{1-x}[(DyHoErNb)O_7]_x$ $(0 \ll x \ll 1)$, denoted $10CCFBO_{xNb}$.[88] (a) Schematic illustration of the long- and short-range order in $10CCFBO_{xNb}$. A long-range order–disorder transition (ODT) occurs at $x$ = 0.81 ± 0.01, from ordered pyrochlore to disordered defect fluorite. Short-range weberite-type order is observed in the long-range disordered fluorite phase and persists across the long-range fluorite-to-pyrochlore ordering transition. (b) Rietveld refinements of neutron diffraction patterns suggest that the ODT occurs through abrupt migration of oxygen anions from the 48*f* to 8*a* sites. (c) Diffuse scattering in neutron total scattering patterns for Nb-rich compositions indicates the presence of short-range weberite-type order. (d) Small-box modeling revealed $C222_1$ weberite-type ordering of the local polyhedral structure at the nanoscale in four compositions near the ODT, which can coexist with long-range pyrochlore order. Replotted from Reference 88 under a CC BY open-access license.

## Ultrahigh-Entropy Ceramics: A Subset of HECs

A series of recent studies, including those discussed above and shown in Figure 5,[88] investigated 10–

21-component fluorite-based CCCs that exhibit intriguing abrupt fluorite–pyrochlore and pyrochlore–weberite phase transitions.[31, 87-89] These 10–21-component CCCs can be considered ultrahigh-entropy ceramics (UECs), which can be defined as ceramic solid solutions with an ideal mixing configurational entropy >2 $k_B$ per atom on at least one sublattice. Based on these definitions, UECs are a subset of HECs.

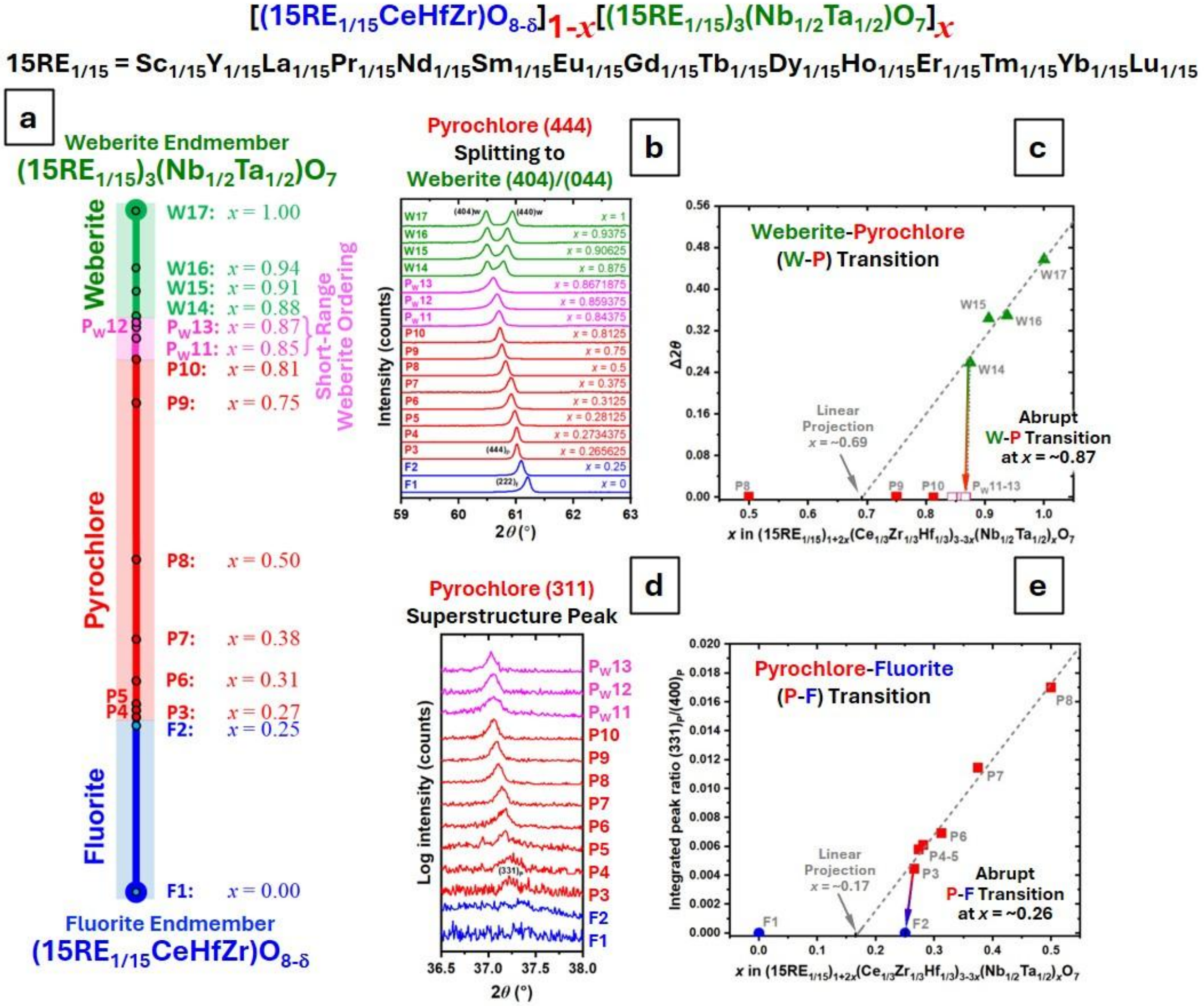

**Figure 6.** Abrupt fluorite-pyrochlore and pyrochlore-weberite phase transitions in a series of 20-component 20CCFBO$_{x\mathrm{Nb/Ta}}$: $[(15\mathrm{RE}_{1/15})_1(\mathrm{Ce}_{1/3}\mathrm{Hf}_{1/3}\mathrm{Zr}_{1/3})_3\mathrm{O}_{7.5}]_{1-x}[(15\mathrm{RE}_{1/15})_3(\mathrm{Nb}_{1/2}\mathrm{Ta}_{1/2})\mathrm{O}_7]_x$, where $15\mathrm{RE}_{1/15}$ refers to the equimolar mixture of 15 rare earth (RE) elements: Sc, Y, La, Pr, Nd, Sm, Eu, Gd, Tb, Dy, Ho, Er, Tm, Yb, and Lu.[89] (a) This series of 20CCFBO$_{x\mathrm{Nb/Ta}}$ all possess single ultrahigh-entropy phases in the cubic $MO_2$ fluorite (F, $0 \leq x \leq 0.25$), cubic $A_2B_2O_7$ pyrochlore (P, $0.2656 \leq x \leq 0.8672$), or orthorhombic $A_3BO_7$ weberite (W, $0.875 \leq x \leq 1$) structure. (b) The pyrochlore-weberite transition is characterized by the splitting pyrochlore (444) peak (using $\Delta 2\theta$ as a weberite order parameter). (d) The fluorite-pyrochlore transition is characterized by the emergence of the pyrochlore (211) super peak (using the normalized integrated peak intensity as the pyrochlore order parameter). With changing compositional variable $x$, this series of 20CCFBO$_{x\mathrm{Nb/Ta}}$ undergoes (e) an abrupt fluorite-pyrochlore transition at $x = \sim 0.27$ and (c) an abrupt pyrochlore-weberite transition at $x = \sim 0.87$. Reprinted from Reference 89 under a CC BY open-access license.

Figure 6 illustrates a series of ultrahigh-entropy, 20-component compositionally complex fluorite-based oxides (20CCFBO$_{x\mathrm{Nb/Ta}}$): $[(15\mathrm{RE}_{1/15})_1(\mathrm{Ce}_{1/3}\mathrm{Hf}_{1/3}\mathrm{Zr}_{1/3})_3\mathrm{O}_{7.5}]_{1-x}[(15\mathrm{RE}_{1/15})_3(\mathrm{Nb}_{1/2}\mathrm{Ta}_{1/2})\mathrm{O}_7]_x$ ($0 \leq x \leq 1$), where $15RE_{1/15}$ denotes an equimolar mixture of 15 rare-earth (RE) elements: Sc, Y, La, Pr, Nd, Sm, Eu, Gd, Tb, Dy, Ho, Er, Tm, Yb, and Lu.[89] Despite the Gibbs phase rule allowing up to 20 phases at thermodynamic equilibrium, 17 of the synthesized $20CCFBO_{x\mathrm{Nb/Ta}}$ compositions form virtually single ultrahigh-entropy phases with fluorite, pyrochlore, or weberite structures, as schematically illustrated in Figure 6a. With increasing $x$, this series undergoes an abrupt fluorite–pyrochlore transition at $x \approx 0.27$ (Figure 6d) and an abrupt pyrochlore–weberite transition at $x \approx 0.87$ (Figure 6b). There are abrupt changes in the order parameters at both phase transitions (Figures 6c and 6e). Notably, weberite-type short-range order can persist into the long-range pyrochlore phase, resulting in the lowest thermal conductivities.

## Dual-Phase CCCs and Thermodynamics

In 2020, Qin *et al.* reported the first dual-phase CCCs, consisting of five-component diboride and monocarbide UHTC phases (Figure 7a).[35] Interestingly, although each specimen contains equimolar amounts of Ti, Zr, Hf, Ta, and Nb, the metal cations partition between the two phases at thermodynamic equilibrium, resulting in non-equimolar cation compositions in both the boride and carbide phases (Figure 7b). Consequently, neither individual phase is "high entropy" according to the conventional definition. A thermodynamic relationship governing the equilibrium compositions of these carbide and boride CCC phases was established, with the measured cation ratios between the boride and carbide phases remaining approximately constant (Figure 7c).[35] Subsequently, a thermodynamic model based on the differential formation energies of borides and carbides calculated by density functional theory (DFT) was developed to explain the observed cation partitioning (Figure 7c).[35] This series of dual-phase CCCs exhibits hardness values exceeding the weighted linear average of the corresponding single-phase high-entropy diboride and carbide counterparts, both of which are already harder than the rule-of-mixtures predictions based on their individual binary constituents.[35] This finding highlights the potential of microstructural engineering in dual-phase CCCs to further enhance materials properties.

Subsequently, various dual-phase CCCs were reported and investigated.[97-102] A thermodynamic relationship was also discovered in 8–14-component bixbyite–fluorite oxide dual-phase CCCs, in which the ratio of the atomic fractions of eight trivalent rare-earth cations between the two equilibrium phases remains approximately constant.[97] Interestingly, this ratio varies linearly with the ionic radius of the rare-earth cations (Figure 7d). Further analysis showed that the differential enthalpy of solution between the bixbyite and fluorite phases approximately follows $\Delta H_i$ (eV) = $-0.80 \times (r_i - 1.06$ Å).[97] In addition, fluorite–pyrochlore dual-phase thermodynamic equilibrium has been investigated.[98] The partitioning of rare-earth (La, Nd, Sm, Eu, Gd, Dy, Ho, Er, Yb, or Lu) cations between the fluorite and pyrochlore phases correlates with cation radius. For rare-earth zirconates, a critical cation radius of ~1.065 Å was identified: rare-earth cations with radii above this value preferentially dissolve in the pyrochlore CCC, whereas those with smaller radii preferentially dissolve in the fluorite CCC. A thermodynamic model based on DFT-calculated fluorite–pyrochlore disorder energies and an ideal-solution model was developed to describe the dual-phase equilibrium. A strong linear correlation was found between the differential free energy of solution of rare-earth cations between the fluorite and pyrochlore phases and the disorder energies of the corresponding individual ternary $RE_2TM_2O_7$ components (Figure 7e).[98] Such thermodynamic relationships provide useful guidance for designing dual-phase CCCs.

In general, dual-phase CCCs provide a paradigm for tailoring functional and mechanical properties through controlled variations in phase fractions, compositions, and microstructures, offering expanded opportunities for microstructural engineering. The individual CCC phases in dual-phase CCCs are intrinsically non-equimolar. Further fundamental studies and thermodynamic modeling are needed to understand and predict dual-phase equilibria in CCCs and enable their rational design.

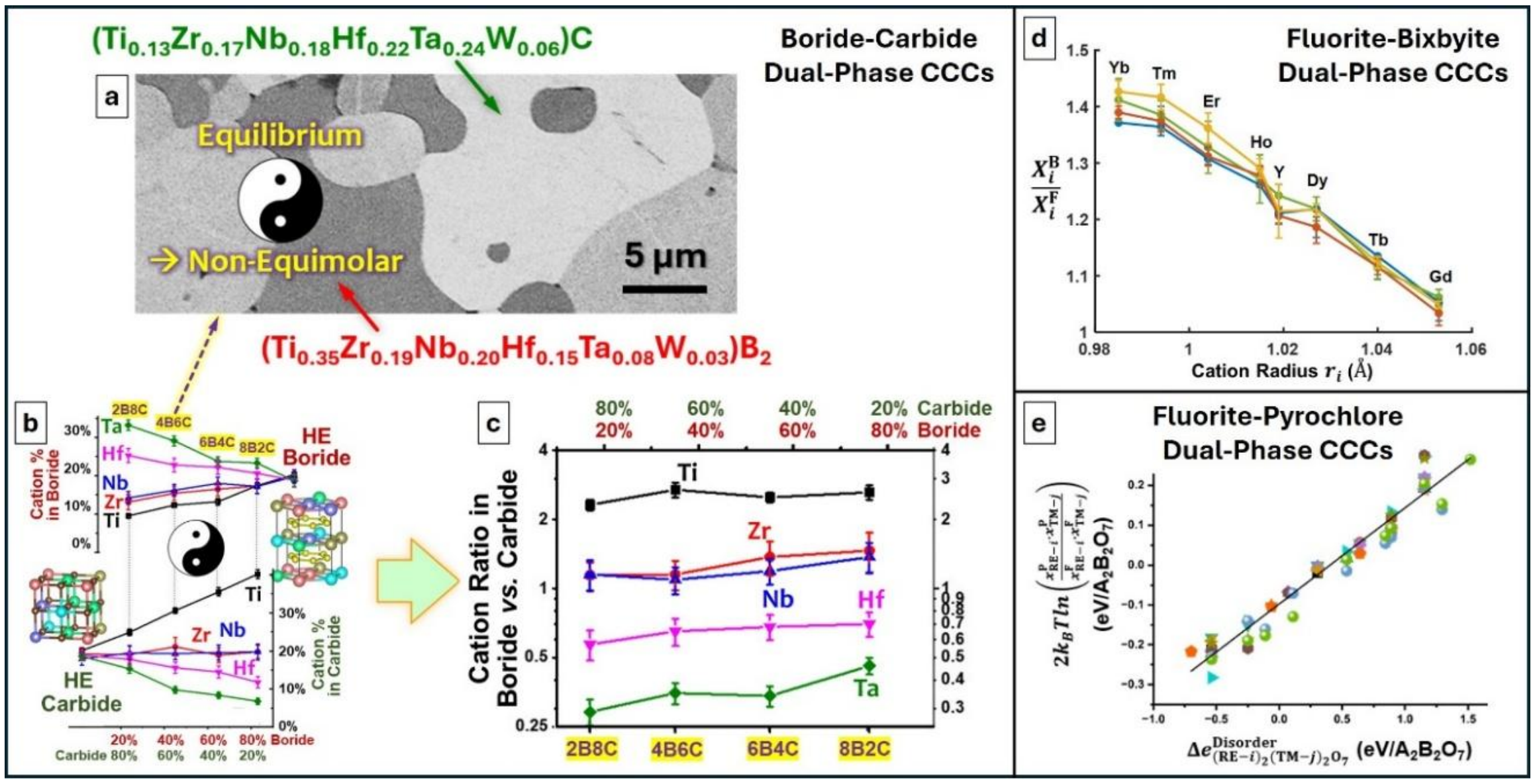

**Figure 7.** In dual-phase CCCs, thermodynamic equilibrium dictates cation partitioning between the two phases; *e.g.*, despite an overall equimolar composition, two non-equimolar CCC phases form at thermodynamic equilibrium. (a-c) Boride–carbide dual-phase CCCs (the first reported dual-phase CCCs and a new class of compositionally complex UHTCs enabling microstructural engineering):[35] (a) Representative scanning electron microscopy (SEM) image of a dual-phase CCC and measured compositions of the boride (diboride) and carbide (rocksalt monocarbide) phases. (b) Measured cation percentages in the boride and carbide phases for a series of dual-phase CCCs with varying nominal phase fractions. (c) Cation ratios between the boride and carbide phases remain approximately constant. A thermodynamic model based on density functional theory (DFT)-calculated formation energies of the borides and carbides is proposed to explain cation partitioning between the two phases. Similar thermodynamic relationships and models can be further extended to oxide dual-phase CCCs. (d) In fluorite–bixbyite dual-phase CCCs, the ratio of the atomic fractions of rare-earth elements in the equilibrium fluorite and bixbyite phases remains approximately constant and varies linearly with the ionic radius of the rare-earth cation.[97] (e) A thermodynamic relationship is identified for the fluorite–pyrochlore dual-phase equilibrium in CCCs, in which the differential free energy of solution between the ordered pyrochlore and disordered fluorite phases is linearly correlated with DFT-calculated disorder energies.[98] Panels (a–c) are adapted with permission from Reference 35, and Panels (d–e) are reprinted from References 97 and 98 under CC BY open-access licenses.

## Interfaces and Microstructural Evolution

As with any other type of materials, microstructures and interfaces (grain and phase boundaries and free surfaces) are key determinants of CCCs, though they remain largely unexplored. The interfacial structures and microstructural evolution in CCCs can be more complex and intriguing in general.

As an example, Figure 8 shows recent studies[74-76] on improving the ionic conductivity of perovskite oxide CCCs through a grain boundary (GB) disordering transition. A recent study developed a new class of compositionally complex perovskite oxide solid electrolytes in which interfacial engineering improves Li-ion conductivity by both promoting grain growth, thereby reducing total GB resistance, and directly increasing specific GB conductivity (Figure 8).[74] Specifically, the ionic conductivity of a new perovskite CCC, $(Li_{0.375}Sr_{0.4375})(Ta_{0.375}Nb_{0.375}Zr_{0.125}Hf_{0.125})O_{3-\delta}$ (LSTNZH), was improved by >60% relative to the state-of-the-art perovskite oxide solid electrolyte ($(Li_{0.375}Sr_{0.4375})(Ta_{0.75}Zr_{0.25})O_{3-\delta}$ (LSTZ) baseline through enhanced grain growth (Figure 8a). Temperature-dependent GB segregation and grain growth in Nb-containing LSTNZH were examined to investigate the underlying mechanisms.[76] Notably, increasing temperature induces significant Nb segregation at GBs (Figure 8b), contrary to classical GB segregation models that predict temperature-induced desorption. This behavior suggests premelting-like GB disordering coupled with Nb segregation, consistent with the relatively low melting temperature of $Nb_2O_5$. This mechanism also explains the observed abnormal grain growth (Figure 8c), which reduces total GB resistance by decreasing the number of GBs. Furthermore, the specific GB ionic conductivity of LSTNZH can be further improved by quenching, which preserves more disordered, Nb-segregated GBs (Figure 8d).[74] The combination of enhanced grain growth and improved specific GB conductivity results in GB-enabled conductivity enhancement in quenched LSTNZH, achieving 2.7× the conductivity of the LSTZ baseline.[74]

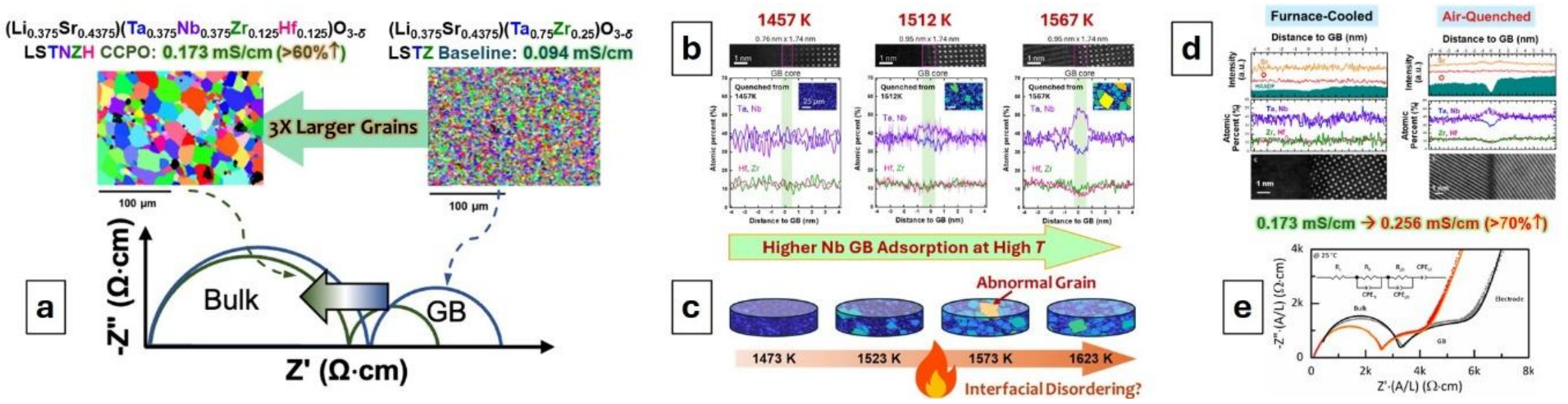

**Figure 8.** An example of using interfacial and microstructural engineering to improve a property (ionic conductivity) of CCCs. In a new class of compositionally complex perovskite oxide solid electrolytes (for solid-state lithium-ion batteries), enhanced Li-ion conductivity is attributed to a grain boundary (GB) transition that promotes grain growth and increases specific GB conductivity.[74] (a) The ionic conductivity of $(Li_{0.375}Sr_{0.4375})(Ta_{0.375}Nb_{0.375}Zr_{0.125}Hf_{0.125})O_{3-\delta}$ is >60% higher than that of the $(Li_{0.375}Sr_{0.4375})(Ta_{0.75}Zr_{0.25})O_{3-\delta}$ (LSTZ) baseline, primarily due to enhanced grain growth that reduces total GB resistance, together with moderately increased specific bulk and GB conductivities. (b) Notably, increasing temperature induces significant GB segregation of Nb, in contrast to classical GB segregation models that predict temperature-induced desorption. This behavior instead suggests premelting-like GB disordering coupled with $Nb_2O_5$ segregation, resulting in (c) accelerated (abnormal) grain growth.[76] (d) Quenched specimens exhibit more disordered GBs with Nb segregation, which further increases specific GB conductivity.[76] (e) Consequently, quenching provides an additional >70% increase in ionic conductivity, reaching 2.7× of the LSTZ baseline.[74] Reprinted from References 74 and 76 under CC BY open-access licenses.

A potentially transformative research direction for interfaces and microstructural evolution in CCCs is to exploit interfacial phase-like (complexion) transitions and construct GB phase diagrams.[103, 104] Another relevant emerging concept is high-entropy grain boundaries (HEGBs),[105] which were introduced to stabilize ultrafine-grained and nanocrystalline alloys at exceptionally high temperatures.[105-108] This concept can be extended to CCCs. As an encouraging example, analogous high-entropy interfaces may help stabilize porous spinel oxide CCCs, enabling ultralow thermal diffusivity (~1000× lower than that of air) and conductivity while maintaining a high elastic modulus and remarkable microstructural stability at 1000 °C.[109]

## Concluding Remarks and Outlook

Over the last decade, the rapid expansion of HECs encompassing diverse chemistries, bonding characters, and crystal structures has revealed vast opportunities. In 2020, we further proposed extending the exploration of HECs to CCCs to reconnoiter compositional complexity across broad non-equimolar spaces, in combination with long- and short-range order, to design CCCs with tailorable and enhanced properties that can outperform their higher-entropy counterparts. Additional opportunities arise from dual-phase CCCs that enable microstructural engineering and from exploiting interfacial phases (also known as "complexions"[104]) in CCCs to control microstructural evolution and material properties. These strategies provide new opportunities to tailor and enhance properties even as configurational entropy decreases, because maximizing entropy is not the ultimate scientific or technological goal.

In general, increasing compositional complexity in CCCs can enhance a range of properties, including ultralow thermal conductivity and superior thermoelectric performance through enhanced phonon scattering; extraordinary high-temperature mechanical properties and improved hardness through impeded dislocation motion; enhanced dielectric energy density in ferroelectric relaxors through increased correlated disorder and reduced polar nanoregion sizes; and boosted catalytic activity through the generation of diverse surface defects. Point defects in CCCs are also scientifically challenging to understand but offer technological opportunities. Because point defects in a given CCC can experience a spectrum of local chemical environments, their formation energies also span a distribution, resulting in statistically distributed defect populations that can be leveraged to tune and enhance properties, such as for solar thermochemical hydrogen generation and chemical looping.[110-115] Enhanced defect clustering and anharmonicity in CCCs may also enable exotic properties, potentially including giant electrostriction and other anomalous coupling properties in fluorite oxide CCCs.[116, 117]

CCCs provide vast compositional spaces for tuning and improving properties while enabling the simultaneous optimization of multiple targeted properties that are otherwise difficult to achieve in simpler ceramics. Here, compositional and microstructural complexity, encompassing non-equimolar compositions, long- and short-range order, aliovalent doping, vacancies, dual-phase microstructural engineering, and interfacial phases, provides a more powerful and versatile framework for materials design than simply maximizing configurational entropy. Realizing the full potential of CCCs requires effective design strategies and predictive models, including AI-driven approaches.

Novel synthesis and processing routes, such as ultrafast sintering (*e.g.*, synthesizing and densifying the most difficult-to-sinter ultrahigh-temperature high-entropy diborides, such as $(Ti_{1/5}Zr_{1/5}Hf_{1/5}Nb_{1/5}Ta_{1/5})B_2$, in merely 2 minutes),[118-120] are critically important and offer benefits in at least three aspects: (1) enabling the otherwise challenging synthesis and processing of complex compositions with controlled homogeneity and microstructures; (2) empowering high-throughput synthesis and processing to support data- and AI-driven materials discovery and exploration across high-dimensional compositional spaces; and (3) attaining non-equilibrium phases, defects, and unique microstructures through far-from-equilibrium processing to achieve unprecedented properties.

### Author contributions

J.L.: Conceptualization, funding acquisition, investigation, resources, visualization, writing—original draft, writing—review and editing.

### Funding

This review article draws on results from various research projects supported by several funding agencies over the past decade, covering different classes of high-entropy and compositionally complex ceramics and their diverse structural and functional properties, including current research supported by the National Science Foundation (NSF) through the Designing Materials to Revolutionize and Engineer our Future (DMREF) program on compositionally complex fluorite oxides for anomalous coupling properties (Grant No. DMR-2522977).

### Conflict of interest

The author declares no conflict of interest.